\documentclass[a4paper,11pt]{article}
\usepackage{pos}
\usepackage{graphicx}
\usepackage{subcaption}
\title{Two-color lattice QCD in $(1+1)$ dimensions with Grassmann tensor renormalization group}
\ShortTitle{Two-color lattice QCD in (1+1)D with Grassmann TRG}

\author*[a]{Kwok Ho Pai}
\author[b,a]{Shinichiro Akiyama}
\author[a,c,d]{Synge Todo}

\affiliation[a]{Department of Physics, The University of Tokyo, 
    Bunkyo-ku, Tokyo 113-0033, Japan}
\affiliation[b]{Center for Computational Sciences, University of Tsukuba, 
    Tsukuba, Ibaraki 305-8577, Japan}
\affiliation[c]{Institute for Physics of Intelligence, The University of Tokyo, 
    Bunkyo-ku, Tokyo 113-0033, Japan}
\affiliation[d]{Institute for Solid State Physics, The University of Tokyo, 
    Kashiwa, Chiba 277-8581, Japan}

\emailAdd{hopai.kwok@phys.s.u-tokyo.ac.jp}
\emailAdd{akiyama@ccs.tsukuba.ac.jp}
\emailAdd{wistaria@phys.s.u-tokyo.ac.jp}    

\abstract{
The $(1+1)$-dimensional two-color lattice QCD is studied with the Grassmann tensor renormalization group. 
We construct tensor network representations of theories with the staggered fermion and the Wilson fermion and show that Grassmann tensor networks can describe both cases with the same bond dimension.
We also propose an efficient initial tensor compression scheme to gauge degrees of freedom.
We compute the number density, chiral condensate, and diquark condensate at finite density, employing the staggered fermions.
For the theory with Wilson fermion, a critical point in the negative mass region is identified by inspecting the pseudoscalar condensate and the conformal field theory data.
}

\FullConference{The 41st International Symposium on Lattice Field Theory (LATTICE2024)\\
 28 July - 3 August 2024\\
Liverpool, UK\\}

\notes{
\note{UTHEP-800, UTCCS-P-164}
}

\begin{document}
\maketitle

\section{Introduction}
The tensor renormalization group (TRG) approach~\cite{Levin:2006jai}, which is a coarse-graining transformation on a tensor network, has attracted a lot of attention in the research of the lattice gauge theory because it does not rely on any sampling procedure and thus free from the sign problem. 
Not only allowing direct estimation of the partition function at finite density or with a topological theta term, one can apply the TRG method to obtain the central charge and scaling dimensions of the underlying conformal field theory (CFT) at some possible criticality. 
In the last decade, TRG has been used to simulate a variety of lattice models in different dimensions. 
The applicability of the method to non-Abelian gauge theories coupled to fermions remains unclear, even in $(1+1)$ dimensions. 
Recently, several TRG studies have explored this direction~\cite{Bloch:2022vqz,Asaduzzaman:2023pyz}.

In this work, we construct a Grassmann tensor network representation for the partition function of $(1+1)$-dimensional two-color lattice QCD and apply the bond-weighted TRG (BTRG) algorithm~\cite{PhysRevB.105.L060402, Akiyama:2022pse} to compute the number density, chiral condensate, and diquark condensate at finite density, employing the staggered fermions.
We also consider the lattice theory with the Wilson fermion and investigate the critical phenomena in the negative mass regime~\cite{Aoki:1983qi}, calculating the pseudoscalar condensate and the eigenvalue spectrum of the transfer matrix.
These investigations demonstrate the efficacy of the TRG in exploring non-perturbative phenomena in lattice gauge theories, paving the way for future studies of the corresponding higher-dimensional models.

\section{Two-color QCD with staggered fermion}

\subsection{Lattice model}
We consider an $SU(2)$ Yang-Mills theory coupled with staggered fermion, defined on a $(1+1)$-dimensional square lattice $\Lambda$ with volume $V=L^2$.
The action $S$ is given by $S = S_f + S_g$ with
\begin{equation} \label{eq: S_f}
    S_f = \sum_{n\in\Lambda,\, \nu=1,2} \frac{p_{\nu}(n)}{2} \left[ \mathrm{e}^{\mu\delta_{\nu,2}} \bar{\chi}(n) U_{\nu}(n) \chi(n+\hat{\nu}) - \mathrm{e}^{-\mu\delta_{\nu,2}} \bar{\chi}(n+\hat{\nu}) U^\dagger_{\nu}(n) \chi(n)\right] + m\sum_n\bar{\chi}(n)\chi(n),
\end{equation}
\begin{equation} \label{eq: S_g}
    S_g = -\frac{\beta}{2} \sum_{n} \mathrm{Re} \mathrm{Tr} U_1(n) U_2(n+\hat{1}) U^\dagger_1(n+\hat{2}) U^\dagger_2(n) ,
\end{equation}
where $n=(n_1, n_2)$ labels the coordinate of each lattice site, and the lattice spacing $a$ has been set to $a=1$. 
The two-component staggered fermions are denoted by $\chi(n)$ and $\bar{\chi}(n)$ with a mass $m$ and chemical potential $\mu$. 
The $SU(2)$ link variable defined on the edge that points from the site $n$ to $n+\hat{\nu}$ is represented by $U_\nu(n)$, and the inverse gauge coupling is $\beta$. 
The staggered phase function $p_\nu(n)$ takes the value $p_1(n)=1$ and $p_2(n)=(-1)^{n_1}$.
The partition function is then given by a Euclidean path integral
\begin{equation} \label{eq: Z}
    Z = \int \mathcal{D}U \mathcal{D}\chi \mathcal{D}\bar{\chi}\,\text{e}^{-S},
\end{equation}
where $\mathcal{D}\chi \mathcal{D}\bar{\chi} \equiv \prod_n \prod_{c=1}^{2} \mathrm{d}\chi_c(n) \mathrm{d}\bar{\chi}_c(n)$, and $\mathcal{D}U \equiv \prod_n {\rm d}U_1(n)\,{\rm d}U_2(n)$ with ${\rm d}U_\nu(n)$ being the Haar measure of $SU(2)$. 
Unless specified otherwise, we take the periodic boundary conditions for link variables and the (anti-)periodic boundary conditions for fermions in the $\hat{1}$ ($\hat{2}$) direction.

\subsection{Tensor network representation} \label{subsection: tn_repr}
We first consider the tensor network representation of the fermion sector $Z_f[U]$:
\begin{align}
\label{eq:zf}
    Z_{f}[U]
    =
    \int \mathcal{D}\chi \mathcal{D}\bar{\chi}~\text{e}^{-S_{f}}.
\end{align}
Following the formalism in Ref.~\cite{Akiyama:2020sfo}, one can rewrite every factor in $~\text{e}^{-S_{f}}$ attributed to the fermion hopping into an integral of some auxiliary Grassmann field, then integrate out the original staggered fermion fields on each lattice site, to obtain a Grassmann tensor network representation of $Z_f[U]$.

Since the partition function in Eq.~(\ref{eq: Z}) is now given by
\begin{align}
\label{eq:Z_full}
    Z=\int \mathcal{D}U Z_{f}[U]~\text{e}^{-S_{g}},
\end{align}
we adopt the discretization scheme in Ref.~\cite{Fukuma:2021cni} which approximates the $SU(2)$ group integration by
\begin{align} 
\label{eq:Umeasure}
    \int {\rm d}U\, f(U) 
    \simeq
    \frac{1}{K} \sum_{i=1}^{K} f(U_i),
\end{align}
with the sample size $K$ and randomly sampled $SU(2)$ matrices $U_i$.
At finite couplings, this results in a tensor network representation of Eq.~(\ref{eq: Z}) with initial bond dimension $16K$. 
In the infinite coupling limit, our representation allows the exact group integration for all link variables on the lattice.
The initial bond dimension, in this case, reduces to $16$ only.
See Ref.~\cite{Pai:2024tip} for more details of our construction.

\subsection{Initial tensor compression}
To accurately recover the integration in Eq.~(\ref{eq:Umeasure}), the sample size $K$ is generally required to be in $O(10)$~\cite{Fukuma:2021cni}.
This implies a huge initial bond dimension in the context of TRG calculations. 
To address this issue, we propose an initial tensor compression scheme that can efficiently reduce the bond dimension of the initial tensors. This is applied before the TRG iterations.

\begin{figure}[htbp]
    \centering
    \includegraphics[scale=0.55]{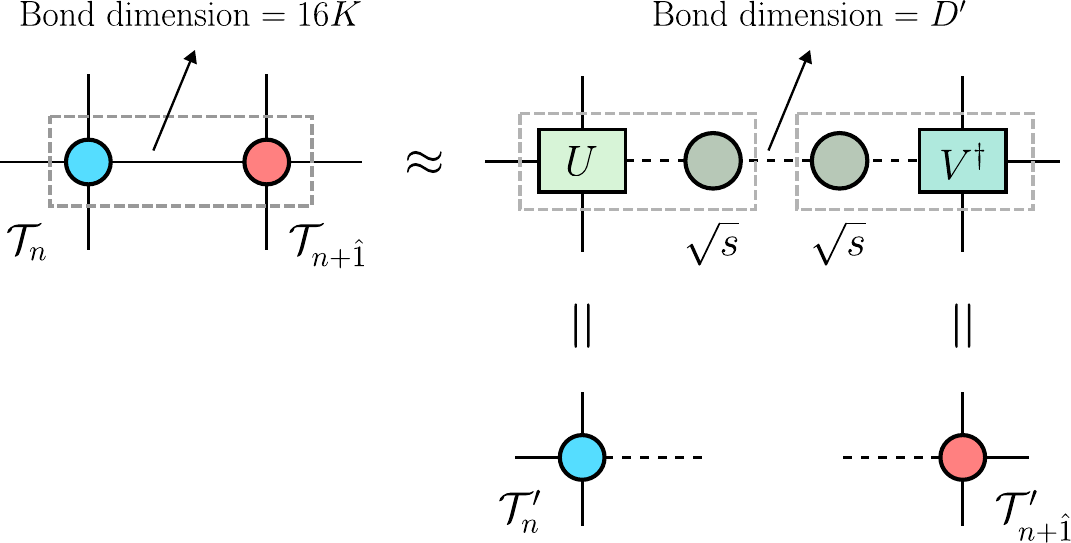}
    \caption{A truncated SVD is performed to reduce the bond dimension of the horizontal link which connects $\mathcal{T}_{n}$ and $\mathcal{T}_{n+\hat{1}}$, from $16K$ to a smaller integer $D'$. Tensors with a reduced bond dimension are denoted as $\mathcal{T}'$.}
    \label{fig:sqz_svd}
\end{figure}

Consider the bond that connects $\mathcal{T}_{n}$ and $\mathcal{T}_{n+\hat{1}}$.
We perform operations that correspond to a truncated singular value decomposition (SVD) on the 6-leg tensor shown in Figure~\ref{fig:sqz_svd}. 
The reduced bond dimension is chosen to be the smallest integer $D'$ that satisfies the following requirement:
\begin{equation} 
    \label{eq:ratio}
    \frac{\sum_{y=1}^{D'} s^2_y}{\sum_{y=1}^{16K} s^2_y} \geq r,
\end{equation}
where $r\leq1$ is a parameter of this compression scheme, and $s$ are the singular values of the truncated SVD with $s_1 \geq s_2 \geq \ldots \geq s_{16K} \geq 0$. 
It turns out that our compression scheme is more efficient in smaller $\beta$. 
As a representative case, the initial bond dimension at $\beta=1.6$ can be reduced to less than half the original number~\cite{Pai:2024tip}.

\subsection{Calculation of observables} \label{subsection: observables}
In this work, we estimate the free energy density $f=\textrm{ln}Z/V$ directly using the BTRG algorithm. 
We would like to compute the expectation value of quark number density $\langle n \rangle$ and chiral condensate $\langle \bar{\chi} \chi \rangle$, which can be written as partial derivatives of $f$:
\begin{equation}
    \label{eq:observables}
    \langle n \rangle = \frac{\partial f}{\partial \mu}, \;\;\;\;\;\; \langle \bar{\chi} \chi \rangle = \frac{\partial f}{\partial m}.
\end{equation}
They are evaluated by the forward difference in this study:
\begin{equation}
    \label{eq:finitedifferecne}
    \langle n \rangle \simeq \frac{f(\mu+\Delta\mu) - f(\mu)}{\Delta\mu} , \;\;\;\;\;\; \langle \bar{\chi} \chi \rangle \simeq \frac{f(m+\Delta m) - f(m)}{\Delta m}.
\end{equation}

In addition, we add a diquark source term to the original action $S$ as
\begin{equation}
    \label{eq: S_diquark}
    S' = S + \frac{\lambda}{2} \sum_{n} \left[\chi^{T}(n) \sigma_2 \chi(n) + \bar{\chi}(n) \sigma_2 \bar{\chi}^T(n) \right]
\end{equation}
with $\lambda \in \mathbb{R}$. 
By Eq.~(\ref{eq: S_diquark}), the expectation value of $\chi \chi \equiv  \sum_{n} \left(\chi^{T} \sigma_2 \chi + \bar{\chi} \sigma_2 \bar{\chi}^T\right)/2V$ can be expressed as
\begin{equation}
    \label{eq: diquark}
\langle \chi \chi \rangle \equiv \frac{1}{2V} \int \mathcal{D}U \mathcal{D}\chi \mathcal{D}\bar{\chi} \, \sum_{n} \left(\chi^{T} \sigma_2 \chi + \bar{\chi} \sigma_2 \bar{\chi}^T\right) \, \textrm{e}^{-S'} = \frac{\partial f}{\partial \lambda}.
\end{equation}
Similarly, $\langle \chi \chi \rangle$ is evaluated by the forward difference:
\begin{equation}
    \label{eq: finite_diquark}
    \langle \chi \chi \rangle \simeq\frac{f(\lambda+\Delta\lambda) - f(\lambda)}{\Delta\lambda}\,\, . 
\end{equation}

\subsection{Numerical results}
The two-color lattice QCD theory with staggered fermion has the $U_V(1) \times U_A(1)$ symmetry at a finite chemical potential $\mu$, in the vanishing $m$ and $\lambda$ limits. 
However, it is well-known that spontaneous breaking of continuous symmetry in two dimensions is not allowed ~\cite{Mermin:1966fe,Coleman:1973ci}. 
We always set $m>0$ and/or $\lambda>0$ in this study, which explicitly breaks the $U_A(1)$ and $U_V(1)$ symmetries, respectively. 
This allows us to check whether the TRG approach can properly reproduce the behavior of physical observables as seen in higher dimensions.

\begin{figure}[ht]
    \captionsetup[subfigure]{justification=centering}
         \subfloat[$m=0.1$ and $\beta=0$]{%
            \includegraphics[width=.5\linewidth]{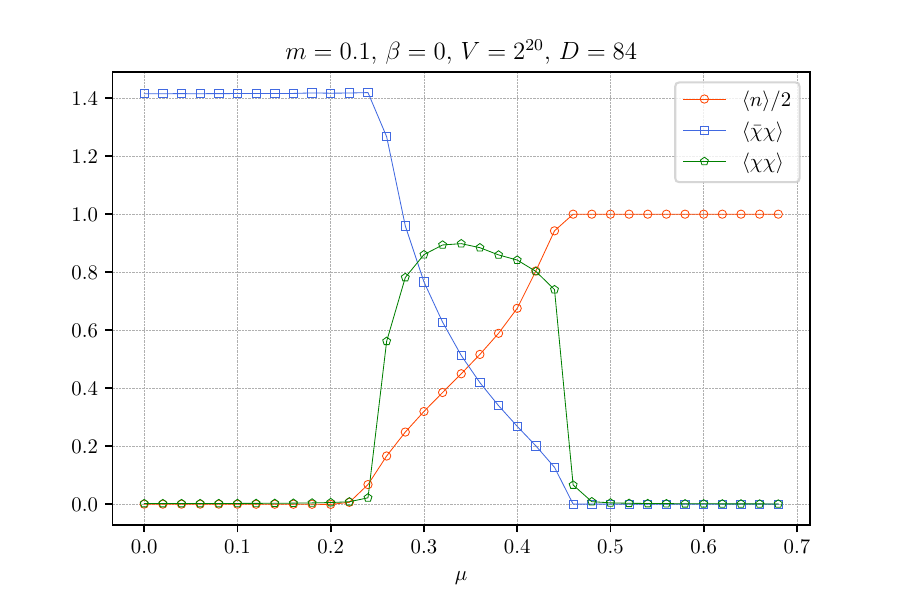}%
            \label{fig:m0.1_beta0}%
        }\hfill
        \subfloat[$m=1$ and $\beta=0$]{%
            \includegraphics[width=.5\linewidth]{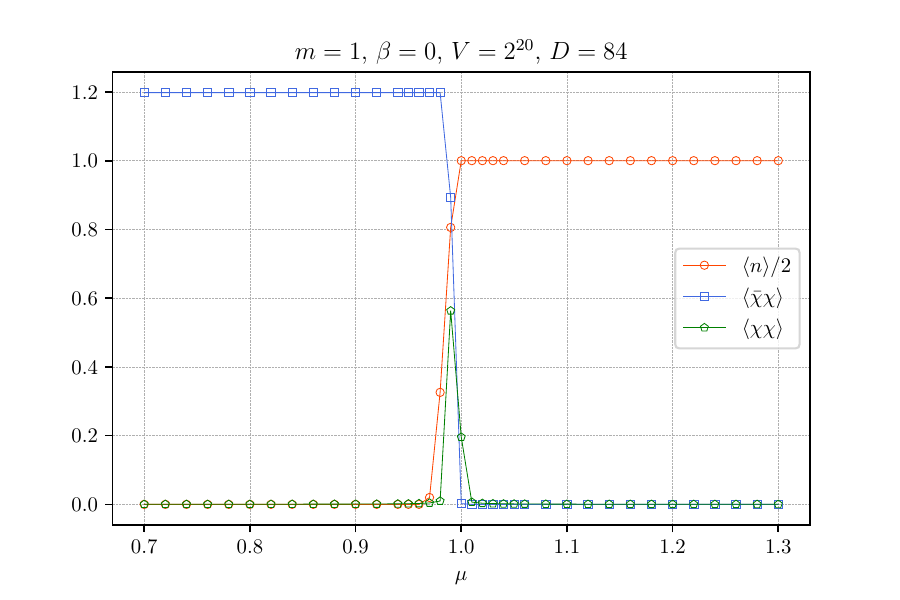}
            \label{fig:m1_beta0}%
        }\\
        \subfloat[$m=0.1$ and $\beta=0.8$]{%
            \includegraphics[width=.5\linewidth]{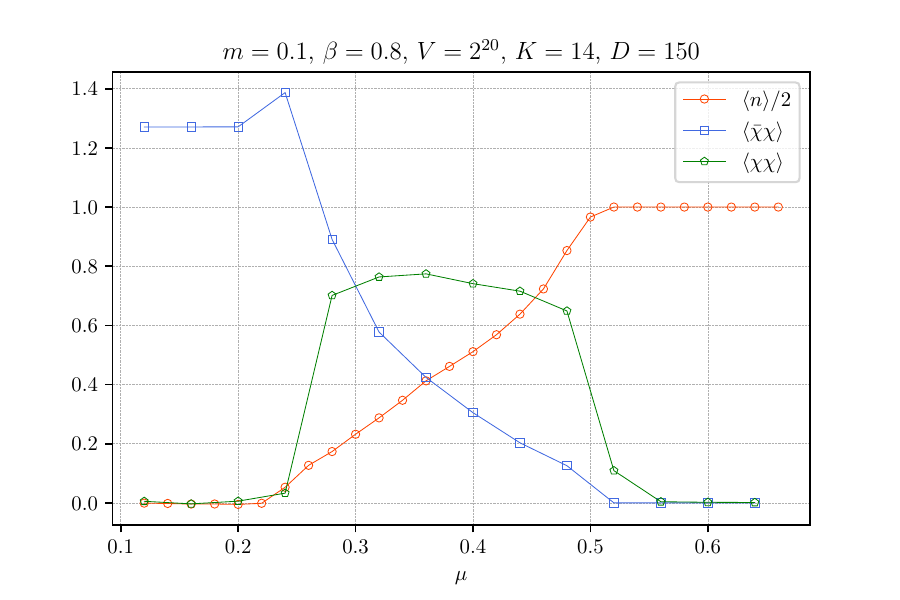}%
            \label{fig:m0.1_beta0.8}%
        }\hfill
        \subfloat[$m=1$ and $\beta=0.8$]{%
            \includegraphics[width=.5\linewidth]{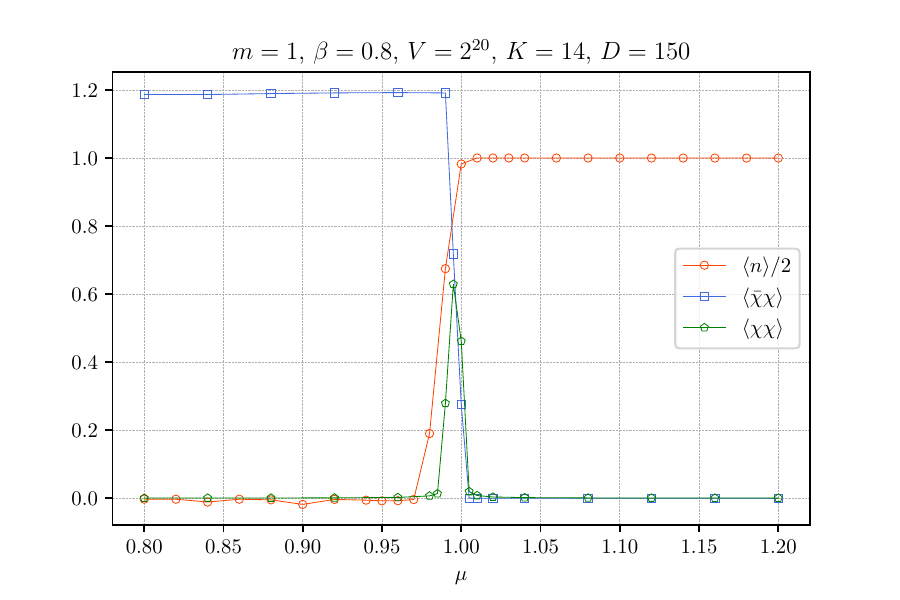}%
            \label{fig:m1_beta0.8}%
        }
        \caption{$\langle n \rangle$, $\langle \bar{\chi} \chi \rangle$, and $\langle \chi \chi \rangle$ as a function of $\mu$ at different $m$ and $\beta$. $D$ refers to the bond dimension cutoff in TRG iterations. To evaluate the numerical differences in Eq.~(\ref{eq:finitedifferecne}) and Eq.~(\ref{eq: finite_diquark}), we set $\Delta \mu=0.04$ for $m=0.1$, $\Delta \mu=0.02$ for $m=1$, $\Delta m = 10^{-4}$, and $\lambda=\Delta\lambda=10^{-4}$. The figures are adapted from Ref.~\cite{Pai:2024tip}.}
        \label{fig:mainresults_staggered}
\end{figure}

In Figure \ref{fig:m0.1_beta0}, $\langle n \rangle$, $\langle \bar{\chi} \chi \rangle$, and $\langle \chi \chi \rangle$ at $m=0.1$, $\beta=0$, and $V=2^{20}$ are shown as a function of $\mu$. 
In the small $\mu$ regime, we observe the Silver-Blaze phenomenon, where the three observables are independent of the chemical potential, up to some point $\mu_{c1} \approx 0.22$. 
When $\mu>\mu_{c1}$, we see an intermediate phase characterized by a finite $\langle \bar{\chi} \chi \rangle$ and $0 < \langle n \rangle < 2$. 
At $m=0.1$, this phase extends over a finite range of chemical potential $\mu_{c1} < \mu < \mu_{c2}$, with $\mu_{c2} \approx 0.46$. 
For $\langle \bar{\chi} \chi \rangle$, it takes a constant finite value in the Silver-Blaze region and decreases in the intermediate phase. 
For $\mu > \mu_{c2}$, $\langle n \rangle$ saturates to two, the maximum allowed by the Pauli exclusion principle on a lattice, while $\langle \bar{\chi} \chi \rangle$ and $\langle \chi \chi \rangle$ drops to some very small values.
We note that the qualitative behavior of these observables is similar to that described in the previous four-dimensional mean-field study~\cite{Nishida:2003uj}.

As illustrated in Figure~\ref{fig:m1_beta0}, a sharper transition occurs with a larger quark mass $m=1$.
The intermediate phase now becomes a very narrow region in $\mu$, with $\mu_{c1}\approx0.96$ and $\mu_{c2}\approx1$. 
Aside from this difference, the qualitative behavior of the three observables at $m=1$, $\beta=0$ is similar to that at $m=0.1$, $\beta=0$.

We also include the results at a finite coupling $\beta=0.8$, with quark masses $m=0.1$ and $m=1$ in Figure~\ref{fig:m0.1_beta0.8} and \ref{fig:m1_beta0.8}, respectively. 
At finite couplings, a discretization of the gauge group integration in Eq.~(\ref{eq:Umeasure}) is required. 
We set the sample size $K=14$ in this study. 
It can be seen that for both choices of mass, the behavior of the observables does not change much as $\beta$ becomes finite.

\begin{figure}[htbp]
    \centering
    \includegraphics[scale=0.7]{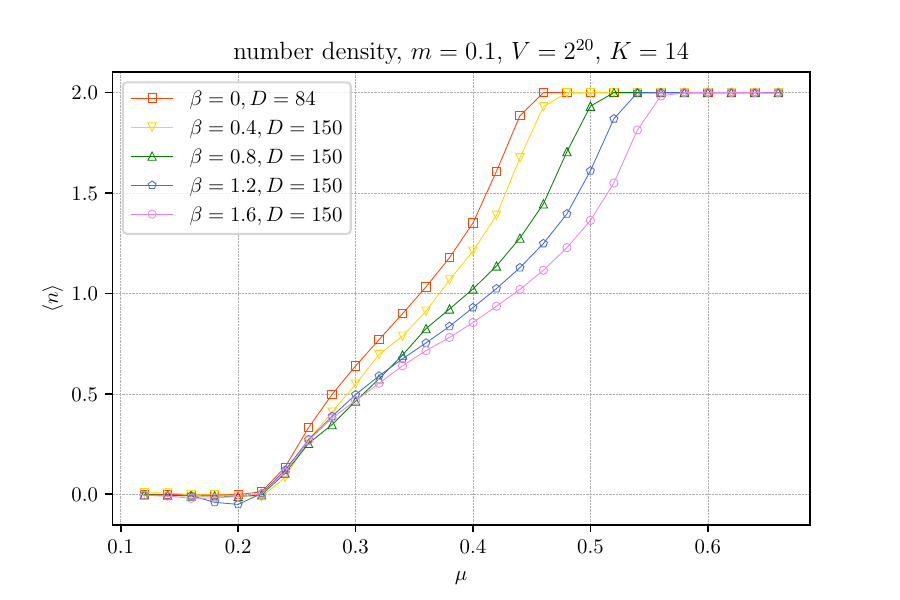}
    \caption{The quark number density $\langle n \rangle$ as a function of chemical potential $\mu$ at $m=0.1$, $\beta=0, 0.4, 0.8, 1.2, 1.6$ in the thermodynamic limit. The bond dimension in the infinite coupling calculation is $D=84$, and the bond dimension in the finite $\beta$ calculations is $D=150$. To evaluate the numerical differences in Eq.~(\ref{eq:finitedifferecne}), we set $\Delta \mu=0.04$. The figure is adapted from Ref.~\cite{Pai:2024tip}.}
    \label{fig:number_betas}
\end{figure}

Finally, we investigate how $\mu_{c1}$ and $\mu_{c2}$ change with $\beta$ at $m=0.1$. 
In Figure~\ref{fig:number_betas}, $\langle n \rangle$ calculated at different $\beta$ are presented as a function of $\mu$. 
We see that $\mu_{c1} \approx 0.22$ for $0\le\beta\le1.6$. 
For $\mu_{c2}$, it is located at larger $\mu$ as $\beta$ increases. 
It is expected because the quarks do not saturate in regions of larger $\mu$ when approaching the continuum limit or weakening the gauge coupling equivalently.

\section{Two-color QCD with Wilson fermion}
In the following, we present some preliminary results on the $(1+1)$-dimensional two-color lattice QCD with one-flavor Wilson fermion in the infinite coupling limit, using the TRG approach.

We consider the following fermion action, instead of Eq.~\eqref{eq: S_f},
\begin{align}
    \label{eqn: S_Wilson}
    S_f 
    &=- \frac{1}{2} \sum_{n,\nu}{ \left[ \bar{\psi}(n) \left(1-\gamma_{\nu}\right) U_{\nu}(n) \psi(n+\hat{\nu}) + \bar{\psi}(n) \left(1+\gamma_{\nu}\right) U^\dagger_{\nu}(n-\hat{\nu}) \psi(n-\hat{\nu}) \right]} \nonumber\\
    &+(m+2r)\sum_{n}{\bar{\psi}(n) \psi(n)}
\end{align}
where $\psi(n)$ and $\bar{\psi}(n)$ in two dimensions are four-component Grassmann fields defined on a lattice site $n$.
In Eq.~\eqref{eqn: S_Wilson}, the chiral symmetry is explicitly broken even with the vanishing $m$ due to the Wilson parameter $r$.
In this work, we use the following gamma matrices: $\gamma_1=\sigma_1$, $\gamma_2=\sigma_3$.
The hopping terms in Eq.~(\ref{eqn: S_Wilson}) will have an identical structure as those in Eq.~(\ref{eq: S_f}) after diagonalizing $(1\pm\gamma_1)$. 
As a result, we can use the same strategy in section~\ref{subsection: tn_repr} to obtain a tensor network representation for the one-flavor Wilson fermion theory with the same initial bond dimension $16$ as in the staggered fermion case.

To detect any spontaneous breaking of parity symmetry as reported in the previous TRG study on the lattice Schwinger model with one-flavor Wilson fermion~\cite{Shimizu:2017onf}, we add the following source term to $S_f$:
\begin{equation}
    \label{eqn: S_h}
    S_h = h \sum_{n}{\bar{\psi}(n)\gamma_5\psi(n)},
\end{equation}
where $h\in\mathbb{R}$, and $\gamma_5=-\mathrm{i}\gamma_1\gamma_2=-\sigma_2$. Now, the partition function is given by
\begin{equation}
    \label{eq: Z_wilson}
    Z = \int \mathcal{D}U \mathcal{D}\psi \mathcal{D}\bar{\psi} ~\text{e}^{-S_{g}-S_{f}-S_h}.
\end{equation}
This allows us to find the pseudoscalar condensate $\langle \bar{\psi} \gamma_5 \psi \rangle$, which is an order parameter to detect the parity symmetry breaking, by $\partial f_W /\partial h$ where $f_W=\mathrm{ln}Z_W/V$.
As in section~\ref{subsection: observables}, we compute this partial derivative using the forward difference.

\subsection{Calculation of CFT data}

Since critical points possess conformal symmetry and can be classified into different universality classes according to the scaling behavior of physical quantities near the criticality, it is useful to determine the underlying conformal field theory at some possible critical point. 

\begin{figure}[htbp]
    \centering
    \includegraphics[scale=0.6]{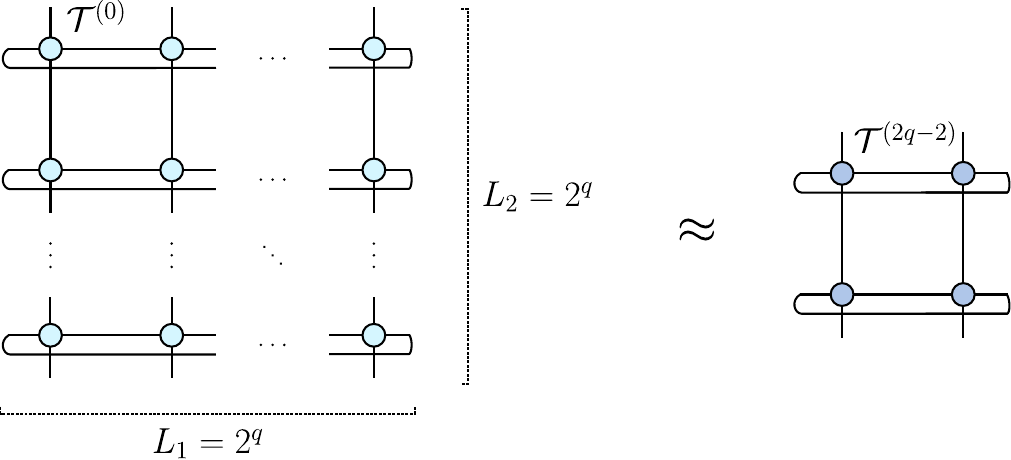}
    \caption{Construction of the transfer matrix for a lattice volume $V=2^{2q}$ by contracting four renormalized tensors $\mathcal{T}^{(2q-2)}$. Here, $\mathcal{T}^{(0)}$ denotes the initial tensor.}
    \label{fig:transfer}
\end{figure}

As suggested in Ref.~\cite{Gu:2009dr}, the transfer matrix of a system in the thermodynamic limit can be formed by contracting some renormalized tensors $\mathcal{T}^{(p)}$ as shown in Figure~\ref{fig:transfer}. 
The superscript $p$ means that the renormalized tensor $\mathcal{T}^{(p)}$ is obtained after $p$ TRG iterations. One should note that a large $p$ limit is implied by the thermodynamic limit.
Then, the central charge $c$ and scaling dimensions $\Delta_i$ at some lattice volume $V=2^{2q}$ can be found by
\begin{align}
    \label{eq: cftdata}
     c = \frac{6}{\pi} \textrm{ln}\left[\lambda_0^{(2q-2)}\right], \; \; \; \; \Delta_i = \frac{1}{2\pi}\textrm{ln}\left[\lambda_0^{(2q-2)} / \lambda_i^{(2q-2)} \right],
\end{align}
respectively, where $\lambda^{(2q-2)}_0\geq\lambda_1^{(2q-2)}\geq\ldots\geq0$ are eigenvalues of the transfer matrix constructed by properly normalized $\mathcal{T}^{(2q-2)}$.
The expression of $c$ in Eq.~(\ref{eq: cftdata}) is only valid in the critical case because it used the scale invariance property $\lambda_0^{(2q-2)}/Z_q=\lambda_0^{(2q-4)}/Z_{q-1}$ with $Z_q$ being the partition function evaluated at a lattice volume $V=2^{2q}$.
For more details of the discussion, one can refer to Ref.~\cite{Ueda:2024ixq} for example.

\subsection{Numerical results}

\begin{figure}[htbp]
        \begin{minipage}[t]{0.48\linewidth}
        \includegraphics[width=\linewidth]{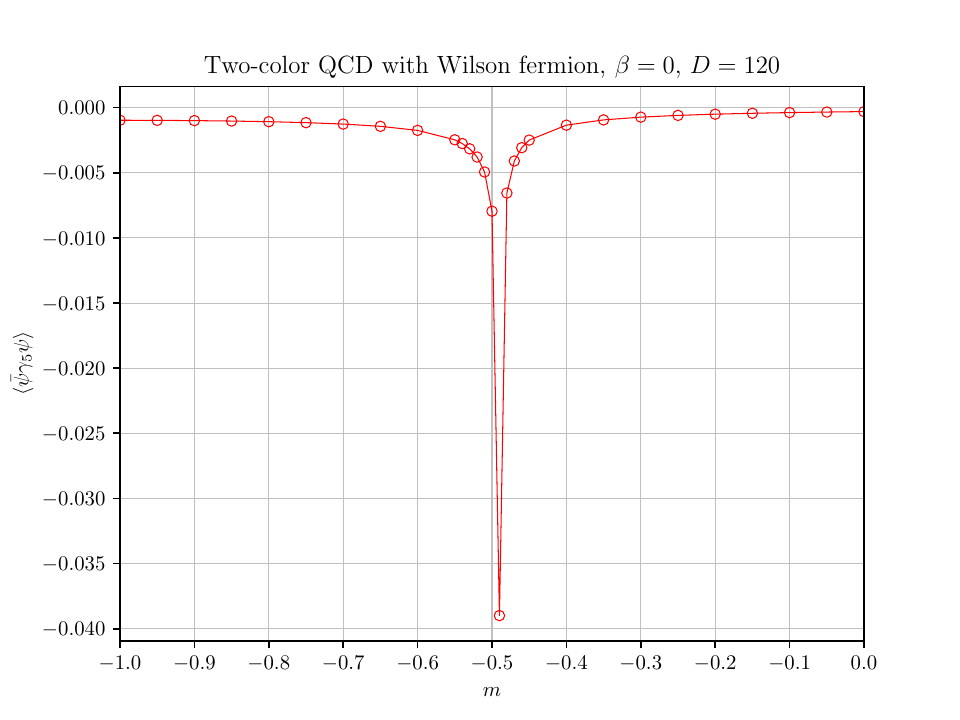}
        \caption{$\langle \bar{\psi} \gamma_5 \psi \rangle$ as a function of $m$ in the infinite coupling limit at a finite $h=10^{-4}$ for the lattice volume $V=2^{20}$. The bond dimension in the calculation is $D=120$. To evaluate the numerical difference, we set $\Delta h=10^{-4}$. }
        \label{fig:psuedoscalar}
        \end{minipage}
    \hfill
        \begin{minipage}[t]{0.48\linewidth}
        \includegraphics[width=\linewidth]{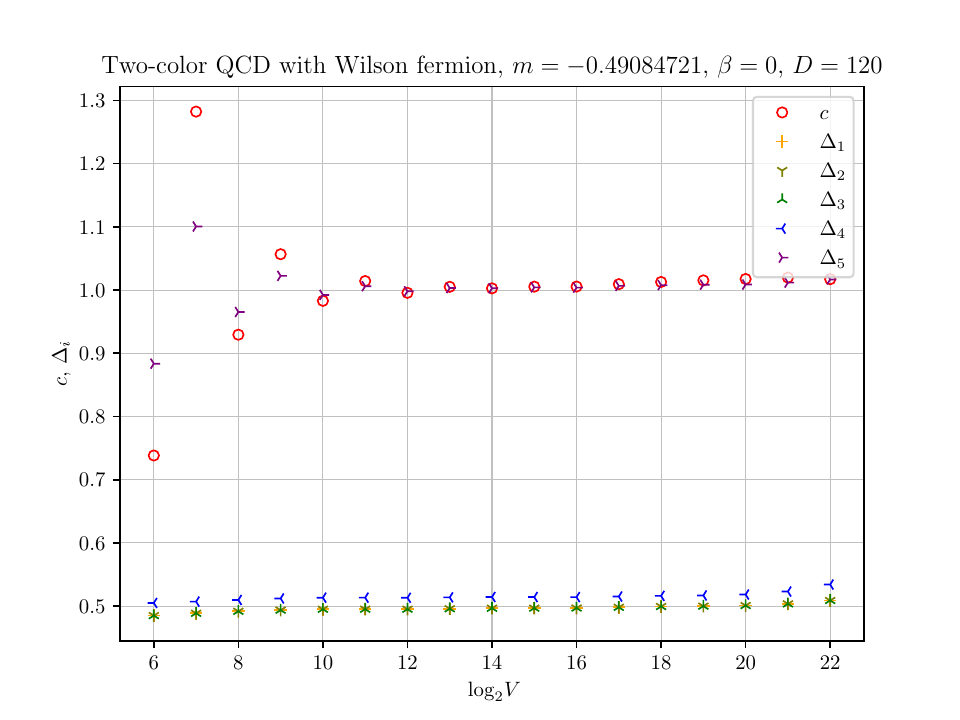}
        \caption{Central charge $c$ and the first five scaling dimensions $\Delta_i$  for different lattice volumes $V$ at $m=-0.49084721$. The bond dimension is $D=120$.}
        \label{fig:cft_infcoupl}
        \end{minipage}%
\end{figure}

Hereafter, we consider the theory in the strong coupling limit $\beta\to0$ and set $r=1$.
Figure~\ref{fig:psuedoscalar} shows the pseudoscalar condensate $\langle \bar{\psi} \gamma_5 \psi \rangle$ as a function of $m$ at a lattice volume $V=2^{20}$, calculated with $h=\Delta h=10^{-4}$.
We observe a peak near $m=-0.5$. 
At $m\approx-0.49084721$, a critical point with $c=1.017453$ at $V=2^{20}$ is spotted.
As shown in Figure \ref{fig:cft_infcoupl}, the first four scaling dimensions at this point are approximately $1/2$ in the thermodynamic limit.
This suggests that the CFT describing this criticality might be the compactified free boson with a radius $R=1/\sqrt{2}$.
Further investigations are in progress to reveal the phase structure of the theory in the negative mass region $-2\leq m\leq0$ and at finite couplings.

\section{Conclusion}
In this study, we construct a Grassmann tensor network representation for the partition function of $(1+1)$-dimensional two-color lattice QCD with the staggered fermion and with the Wilson fermion. 
We propose an efficient compression scheme for the initial tensors, which have a large bond dimension due to the enormous internal degrees of freedom of the theory.
We apply the bond-weighted TRG algorithm to contract the tensor network and compute physical quantities of interest, particularly at finite density.
For the staggered fermion case, we explicitly break the global symmetries of the lattice theory and observe a similar phase structure as in the four dimensions where spontaneous symmetry breaking occurs.
For the Wilson fermion, we are trying to study the critical behavior in the negative mass regime, and some preliminary results at the infinite coupling limit are shown.

These results encourage the further exploration of two-color QCD in various parameter regions and suggest that the TRG approach can provide valuable insights into the non-perturbative dynamics of lattice gauge theories.
Furthermore, investigating the CFT data in the case of the theory with the staggered fermion would be an intriguing direction for future study.

\acknowledgments

A part of the numerical calculation for the present work was carried out with Ohtaka provided by the Institute for Solid State Physics, the University of Tokyo. 
This work is supported by the Endowed Project for Quantum Software Research and Education, the University of Tokyo (\url{https://qsw.phys.s.u-tokyo.ac.jp/}), and the Center of Innovations for Sustainable Quantum AI (JST Grant Number JPMJPF2221).
SA acknowledges the support from JSPS KAKENHI (JP23K13096, JP24H00214) and the Top Runners in Strategy of Transborder Advanced Researches (TRiSTAR) program conducted as the Strategic Professional Development Program for Young Researchers by the MEXT.

\bibliographystyle{JHEP.bst}
\bibliography{bib/for_this_paper,bib/formulation,bib/grassmann,bib/gauge,bib/algorithm,bib/continuous}

\end{document}